# Spatiotemporal profile of emission from oscillating dc microdischarges

T. Kuschel, I. Stefanović, N. Škoro, D. Marić, G. Malović, J. Winter, Z. Lj. Petrović

*Abstract*—The axial light distributions in parallel-plate dc microdischarges in argon show similar behavior to large scale discharges. Between the low-current Townsend mode and the high current glow mode exists a large region of currents where different oscillations appear and the dynamic Volt-Ampere characteristic shows hysteresis behavior. During the oscillations the maximum peak intensity moves closer to the cathode, which is characteristic for the abnormal glow regime even though the average current is considerably smaller.

MICRODISCHARGES have become in the focus of research recently [1,2]. However, only a few studies with parallel plate microdischarges have been made [3]. To check for similarities and breakdown conditions in the microdischarges the Townsend's theory, Paschen law and Volt-Ampere (*V-A*) characteristics could be used [4]. Additionally, axial and radial light emission profiles of microdischarges are essential in order to understand the behavior of discharges in transient regimes [5].

In this work we analyze the axial emission profiles of a 1 mm discharge gap recorded by an ICCD camera in a parallel-plate dc miscrodischarge with 8 mm electrode diameter. The discharge was operated in continuous flow of argon to minimize the influence of impurities on the discharge condition. The *pd* (*p* - pressure, *d* - electrode distance) was kept at 1 Torr×cm (*p* = 10 Torr, *d* = 1 mm), which is the value corresponding to Paschen minimum for argon. The high voltage (HV) pulse (amplitude from volts to 2 kV and duration less than a few ms) was applied on the electrode plates by using loading resistance. The pulsing technique was already used to avoid changes on the electrode surfaces exposed to high-current densities (for detailed description see [4]). By using different HV pulses and loading resistances we were able to stroke the discharge and keep it running in different current modes: low-current diffuse (Townsend), subnormal glow (self-pulsing regime), normal and abnormal glow.

In figure 1, the stars show a static *V-A* characteristic of the microdischarge, where the discharge operates under steady-state conditions. Between the low current and the high current region exists a large gap where current oscillations (self-pulsing) occur, which is represented by characteristic hysteresis. In the following we distinguish between two types of oscillations: i) during relaxation oscillations (label 1) the current develops from almost zero to high currents (several 100 microampere) and ii) during free running oscillations (label 2 and 3) the current oscillates around a high current value (several 100 microampere). Dynamic *V-A* characteristics describes discharge oscillations: the discharge current starts from low current regime and runs through the upper branch to the maximum current value and turns again to the low current. The characteristics during oscillations have circular shapes superimposed with large fluctuations for low currents.

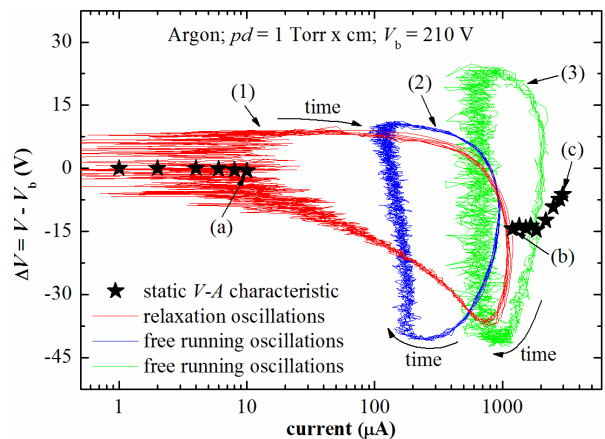

Fig. 1. *V-A* characteristic of the microdischarge: static (stars) and dynamic (color lines). Different labels correspond to the 2D images in figure 2. Static *V-A*: (a) 10 μA, (b) 1 mA, and (c) 2 mA. Dynamic *V-A* during the different oscillations modes: (1) relaxation (2) and (3) free running oscillations. The discharge voltage *V* is normalized to the breakdown voltage $V_b$ to make the small change of voltage upon the large breakdown voltage visible.

Figure 2 shows axial profiles of emission recorded during different high voltage pulses. The upper row corresponds to points on the static *V-A* characteristic indicated in figure 1: (a) Townsend regime, (b) normal glow and (c) abnormal glow. In the Townsend regime (a) the emission increases exponentially from the cathode to the anode; this is characteristic of electron

Manuscript received 1 December 2010; revised 18 May 2011.
T. Kuschel, I. Stefanović and J. Winter are with the Institute for Experimental Physics II, Ruhr-Universität Bochum, D-44780 Bochum, Germany. N. Škoro, D. Marić, G. Malović and Z. Lj. Petrović are with the Institute of Physics, 11080 Belgrade, Serbia.
Work was supported by the Ministry of Science and Technological Development, Serbia (141025), DFG FOR1123, DAAD 50430276, Research Department Plasmas with Complex Interactions.
Publisher Identifier S XXXX-XXXXXXX-X

avalanche in a constant electric field. At the same time, the radial emission profile is symmetric and has a shape of a Bessel function. In the glow regime (b) a peak forms far from the anode marking the formation of the cathode fall. The discharge is constricted and moved to the electrode edge, which is characteristic of the normal glow. By current increase, the discharge switches to the abnormal glow, occupies the whole electrode surface again and the maximum emission moves further to the cathode (c).

The time integrated 2D emission profiles of current oscillations for different voltage pulses are presented in the lower row of figure 2. Surprisingly, the width of the cathode fall during transients (1)-(3) is smaller than that in the normal glow under steady-state conditions (b) and similar to the steady-state conditions of the abnormal glow (c). The radial profile is broadened and more symmetric, which may be expected only for higher current abnormal glow operation.

REFERENCES

[1] K. H. Becker, K. H. Schoenbach and J. G. Eden *J. Phys. D: Appl. Phys.* **39** R55-R70 (2006)
[2] J. G. Eden and S.-J. Park *Plasma Phys. Control. Fusion* **47** B83-B92 (2005)
[3] Z. Lj. Petrović, N. Škoro, D. Marić, C. M. O. Mahony, P. D. Maguire, M Radmilović-Rađenović and G. Malović *J. Phys. D: Appl. Phys.* **41** 194002 (2008)
[4] A. V. Phelps, Z. Lj. Petrović and B. M. Jelenković *Phys. Rev. E* **47** pp. 2825-2838 (1993)
[5] D. Marić, G. Malović and Z. Lj. Petrović *Plasma Sources Sci. Technol.* **18** 034009 (2009)

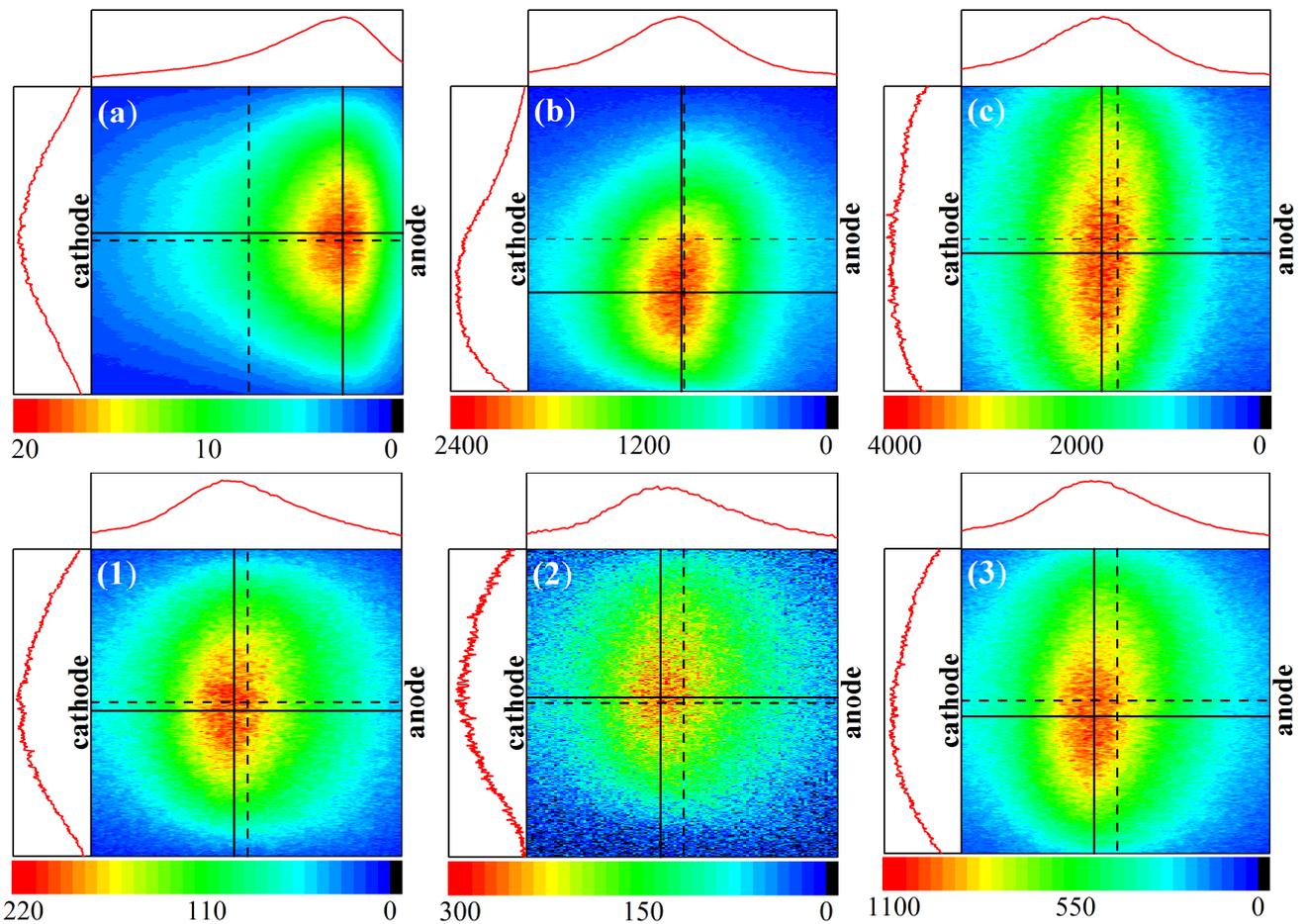

Fig. 2. Time integrated 2D images of the parallel-plate dc microdischarge in argon. The vertical axis shows in radial direction, the horizontal in axial direction and is artificially enlarged to stress the changes in the axial light distribution. Black dashed lines indicate the central axes of the discharge chamber. The graphs on top and on the left side of each image show the emission along the corresponding axis through the peak of emission as indicated by solid black lines. The upper row corresponds to the different discharge regimes on the *static V-A* characteristics in figure 1: (a) Townsend regime, (b) normal glow and (c) abnormal glow. The lower row presents the light distributions time integrated over several oscillations periods for different amplitude of HV pulses: (1) relaxation oscillations, (2) and (3) free running oscillations. The intensity is in arbitrary units.